\documentclass[conference,letterpaper]{IEEEtran}

\usepackage[utf8]{inputenc} 
\usepackage[T1]{fontenc}
\usepackage{url}
\usepackage{ifthen}
\usepackage{cite}
\usepackage{amssymb}
\usepackage{dsfont}
\usepackage[linesnumbered, ruled]{algorithm2e}
\usepackage{graphicx}
\usepackage[cmex10]{amsmath} 

\newcommand{\di}[1]{\Delta I_{#1}}
\newcommand{\irange}{\mathcal{I}}
\newcommand{\leastidx}{L_L}
\newcommand{\greatestidx}{L_U}

\begin{document}
\title{Efficiently Computable Converses for Finite-Blocklength Communication} 


\author{
  \IEEEauthorblockN{Felipe Areces, Dan Song, and Richard Wesel}
  \IEEEauthorblockA{Dept. of Electrical and Computer Engineering \\
  University of California, Los Angeles\\
                    Los Angeles, CA USA\\
                    Email: \{fareces99,dansong,wesel\}@ucla.edu}
  \and
  \IEEEauthorblockN{Aaron B.~Wagner}
  \IEEEauthorblockA{School of Electrical and Computer Engineering\\
                    Cornell University\\ 
                    Ithaca, NY 14853 USA \\
                    Email: wagner@cornell.edu}
}

\maketitle

\begin{abstract}
This paper presents a method for computing a finite-blocklength converse for the rate of fixed-length codes with feedback used on discrete memoryless channels (DMCs). The new converse is expressed in terms of a stochastic control problem whose solution can be efficiently computed using dynamic programming and Fourier methods. For channels such as the binary symmetric channel (BSC) and binary erasure channel (BEC), the accuracy of the proposed converse is similar to that of existing special-purpose converse bounds, but the new converse technique can be applied to arbitrary DMCs. We provide example applications of the new converse technique to the binary asymmetric channel (BAC) and the quantized amplitude-constrained AWGN channel.
\end{abstract}

{\let\thefootnote\relax\footnote{{This research is supported by National Science Foundation (NSF) grant CCF-1955660. Any opinions, findings, and conclusions or recommendations expressed in this material are those of the author(s) and do not necessarily reflect views of the NSF.}}}

\section{Introduction}\label{intro}

Consider communicating one of $M$ messages using a finite blocklength transmission of $n$ symbols over a discrete memoryless channel (DMC) with average error probability $\epsilon$. A fundamental practical problem that arises from this setup is the computation of upper bounds on the size of $M$ (converse) in the finite-blocklength regime when both $n$ and $\epsilon$ are fixed. A hypothesis testing framework can be used to establish converse bounds for arbitrary DMCs \cite{Blahut,Altug:SP:Asymmetric,Altug:EA:Symmetric,ppv_2010}. However, these bounds are generally hard to compute at finite blocklengths due to the high dimensionality of their parameter space. Their computation can be significantly reduced by exploiting symmetry in some cases, such as that of the binary-symmetric channel (BSC), the binary-erasure channel (BEC)~\cite{polyanskiy_saddlepoint}, and the AWGN channel \cite{ErsegheAWGN}. Their limits in various asymptotic regimes can also be computed. These insights provides no assistance at finite blocklengths for arbitrary DMCs, however.

Recent advances toward this goal include an algorithm for saddle point identification \cite{elkayam2016calculation} in the ``minimax'' version of the bound~\cite{polyanskiy_saddlepoint} that is polynomial-time in the blocklength for DMCs, a linear programming formulation of the bound \cite{matthews_linear_2012} that is also polynomial-time for DMCs, and a method for generating good output distributions for general minimax converses where an optimization over a distribution on the channel output is required \cite{kosut_boosting_2015}. The complexity of these approaches is exponential in the input alphabet however, making them unsuitable for even moderately sized channels at longer blocklengths. See \cite{vilar2018saddlepoint} and \cite{ErsegheFull} for approximations of the bound that are more easily computed.

Stochastic control methods have proven useful in deriving results for communication systems with feedback, such as a general framework for the computation of channel capacity \cite{tatikonda_capacity_2009} and the timid/bold technique to improve second-order coding performance \cite{TimidBold}. In many circumstances, stochastic control problems require the design of an optimal controller and dynamic programming (DP) techniques are often used to solve this problem efficiently \cite{tatikonda_control_2000}.

\subsection {Contributions and organization}
In this paper, we develop an efficient algorithm to compute converses at finite blocklength for arbitrary DMCs.
The key idea is to consider the channel with feedback, which allows for efficient computation of the resulting hypothesis
testing bounds using dynamic programming and Fourier methods. We describe this algorithm, analyze its complexity,
and demonstrate its utility on several channels. For the BSC and BEC, we find that it provides bounds that
are close to those obtained via channel-specific methods. On the other hand, we also show that our bound scales well to DMCs with
moderately large alphabets at moderate blocklengths, for which there is currently no known way of efficiently computing nontrivial converse bounds.
Of course, any converse bound for a DMC with feedback also applies to the channel without feedback. 



The rest of the paper is organized as follows. In Section \ref{converses} we formulate the converse bound as a DP problem. In Section \ref{algorithms} we present an algorithm to compute the bound an analyze its complexity. Section \ref{results} presents numerical results for the BSC, BEC, BAC, and quantized amplitude constrained AWGN channel, and provides a comparison to existing bounds where possible. If no specific bounds are available we use Lemma 14 in \cite{TimidBold} with a capacity achieving distribution to obtain a general achievability result without feedback.

\section{DMC Converses via Dynamic Programming} \label{converses}
\subsection{Definitions}
\subsubsection{Probability of Success/Failure}\label{DefPSF}

Assume a DMC with finite input alphabet $\mathcal{X}$ and finite output alphabet $\mathcal{Y}$. In this case, for an $(n,R)$ code with ideal feedback and error probability $\epsilon\in(0,1)$ we define:
\begin{align}\label{def_mfb}
    M^*(n,\epsilon):=\max\{\lceil\exp(nR)\rceil\in\mathbb{N}_+:
    \bar{\textbf{P}}_{\text{e,fb}}(n,R)\leq \epsilon\}\,,
\end{align}
where $\bar{\textbf{P}}_{\text{e,fb}}(n,R)$ is the minimum average error probability achievable by any $(n,R)$ code with feedback. We follow~\cite{TimidBold}. In particular, given input and output random vectors 
\begin{align}
    \textbf{X}^n &= (X_1, \dots, X_n) \in \mathcal{X}^n \\
    \textbf{Y}^n &= (Y_1, \dots, Y_n) \in \mathcal{Y}^n\,,
\end{align}
a (stochastic) controller 
\begin{align}
    F & = F^n\\
    F^k &= (F_1,\dots, F_k)\\
    F_t &: (\textbf{X}^{t-1},\textbf{Y}^{t-1})\mapsto X_{t}\,,
\end{align}
a channel $W\in \mathcal{P}(\mathcal{Y}|\mathcal{X})$, an information density threshold $T$, and an arbitrary output distribution $Q\in\mathcal{P}(\mathcal{Y})$, we refer to the event in which the AID at the end of the block exceeds $T$ as a \emph{success}, $S$. The probability of success for any controller can be defined in terms of $\textbf{X}^n$ and $\textbf{Y}^n$ as
\begin{align}\label{success_event_v2}
    P(S_{F,Q,T})=(F \circ W)\left(\log_2\left[\frac{W(\textbf{Y}^n|\textbf{X}^n)}{Q(\textbf{Y}^n)}\right]>T\right)\,,
\end{align}
where $F\circ W$ denotes the joint distribution over $\textbf{X}^n$ and $\textbf{Y}^n$ induced by the
controller and the channel.
Similarly, the probability of failure is defined as $P(\bar{S}_{F,Q,T})=1-P(S_{F,Q,T})$.

\subsubsection{Information density}\label{DefID}
Given an input distribution to a DMC $G\in\mathcal{P}(\mathcal{X})$ and an output distribution $Q\in\mathcal{P}(\mathcal{Y})$ we can define the information density change from one transmission ($\di{}$) as
\begin{align}\label{DeltaI}
    \di{G}=\log_2\left(\frac{W(Y|X)}{Q(Y)}\right)
\end{align}
with range $\irange_G$ and PMF $f_G(\Delta i)$
\begin{align}\label{ProbDeltaI}
    f_G(\Delta i)=\sum_{x\in\mathcal{X}}\sum_{y\in\mathcal{Y}}\mathds{1}\left(\Delta i=\log_2\left[\frac{W(y|x)}{Q(y)}\right]\right)W(y|x)G(x)\,.
\end{align}
We finally define the set of all possible AID changes in a single transmission as
\begin{align}
    \irange=\left\{\log_2\left(\frac{W(y|x)}{Q(x)}\right)\bigg|x\in\mathcal{X},y\in\mathcal{Y}\right\}
\end{align}
and the set of all possible AID levels attainable by any controller at timestep $k$ as
\begin{align}
    \Lambda_k=\left\{\sum_{\gamma\in\irange}\alpha_{\gamma}\gamma\bigg| \alpha_\gamma\in\mathbb{Z}^+\text{ and } \sum_{\gamma\in\irange}\alpha_{\gamma}=k\right\}\, ,
\end{align}
where $\mathbb{Z}^+$ is the set of nonegative integers.

\subsection{Mathematical formulation}
\subsubsection{General setup}\label{GS}
From Lemma 15 in \cite{TimidBold} we know that given parameters $W,n,\rho>0,\epsilon>0$ and $T=\log_2(\rho)$,
\begin{align}\label{TB_Converse}
    R^*_b(n,\epsilon)\leq \sup_{F}\inf_{Q}\frac{1}{n}\left[ T-\log_2\left(\left[P(\bar{S}_{F,Q,T})-\epsilon\right]^+\right)\right]\,,
\end{align}
where $R^*_b(n,\epsilon)$ is the best achievable code rate in bits for the given blocklength $n$ and error probability $\epsilon$. 

\subsubsection{Computation of maximum probability of success}\label{TheoCompPSF}
From the supremum in (\ref{TB_Converse}) we see that we are interested in minimizing $P(\bar{S}_{F,Q,T})$ or, equivalently, maximizing $P(S_{F,Q,T})$. In order to translate this problem into a dynamic programming framework we will focus on the probability of success of the optimal controller at timestep $k$ and a specific AID level, $S^*_k(\Gamma)$, where references to $Q$ and $T$ are omitted for simplicity. With this setup every $S^*_k(\Gamma)$ can be defined in terms of the following recurrence relation as long as the chosen $Q$ is a product distribution:
\begin{align}
\label{TrivDef}
S^*_n(\Gamma)&=\mathds{1}\left(\Gamma\geq T\right)\\
\label{RecRel1}
    S^*_k(\Gamma)&=\max_{G\in\mathcal{P}(\mathcal{X})}\sum_{\Delta i\in\irange_{G}}S^*_{k+1}(\Gamma+\Delta i)f_G(\Delta i),\,k<n\,.
\end{align}
However, we know that the cardinality of the set of all possible achievable AID levels grows polynomially with the blocklength, which makes direct computation infeasible even for moderate values of $n$. In order to address this problem we will quantize the AID in a sequence of $N$ bins $\beta_{i}=(\tau_{i}^L,\tau_{i}^U]$, $i\in \{\leastidx, \leastidx+1, \dots, \greatestidx\}$ of size $\delta$ such that
\begin{align}
    \frac{\tau_{\greatestidx}^U}{n}=\di{{max}}&=\max_{(x,y)\in\mathcal{X}\times\mathcal{Y}}\log_2\left(\frac{W(y|x)}{Q(y)}\right)\\
    \frac{\tau_{\leastidx}^L}{n}=\di{{min}}&=\min_{(x,y)\in\mathcal{X}\times\mathcal{Y}:W(y|x) > 0}\log_2\left(\frac{W(y|x)}{Q(y)}\right)\\
    N&=\left\lceil \frac{\di{{max}}-\di{{min}}}{\delta}\right\rceil\,.
\end{align}
This definition guarantees that all achievable AID levels will be comprised in this range as no single walk can exceed the bin limits. Our scheme also rounds up any AID realization to the upper bound of its corresponding bin (denoted as $\lceil t\rceil_{\delta} = \delta \left\lceil\frac{t}{\delta}\right\rceil$). Equations (\ref{TrivDef}) and (\ref{RecRel1}) thus become
\begin{align}
\hat{S}^*_n(\Gamma)&=\mathds{1}\left(\Gamma\geq T\right)\\
    \hat{S}^*_k(\Gamma)&=\max_{G\in\mathcal{P}(\mathcal{X})}\sum_{\Delta i\in\irange{x}}\hat{S}^*_{k+1}(\Gamma+\lceil\Delta i\rceil_\delta)f_G(\Delta i)
\end{align}
for $\Gamma\in\{\tau^U_{\leastidx},\tau^U_{\leastidx+1},\cdots,\tau^U_{\greatestidx}\}$. Denote the AID random variable
\begin{equation}\label{DeltaIk}
    \di{k} = \log_2\left(\frac{W(Y_k|X_k)}{Q(Y_k)}\right)\,,
\end{equation}
which has distribution depending on the controller $F^k$. We can now show that this rounding strategy provides an upper bound on the probability of success of the optimal controller and thus preserves the validity of the converse
\begingroup
\begin{align}
    P&(S_{F^*,Q,T})
    = \sup_{F} (F\circ W)\left(\sum_{j=1}^n  \di{j} \geq T\right)\\
    \begin{split}
    =& \sup_{F^{n-1}, F_n}\sum_{\Gamma_{n-1}\in\Lambda_{n-1}} \Biggl[ (F^{n-1}\circ W)\left(\sum_{j=1}^{n-1}  \di{j} = \Gamma_{n-1}\right)
    \\&\cdot (F_{n}\circ W)\left( \di{n} \geq T - \Gamma_{n-1}\middle|\sum_{j=1}^{n-1}  \di{j} = \Gamma_{n-1}\right)\Biggr]
    \end{split}\\
    \begin{split}
    =& \sup_{F^{n-1}}\sum_{\Gamma_{n-1}\in\Lambda_{n-1}} \Biggl[ (F^{n-1}\circ W)\left(\sum_{j=1}^{n-1}  \di{j} = \Gamma_{n-1}\right) \\&\cdot \sup_{F_n}(F_{n}\circ W)\left( \di{n} \geq T - \Gamma_{n-1}\right)\Biggr]
    \end{split}\\
    \begin{split}
    \leq & \sup_{F^{n-1}}\sum_{\Gamma_{n-1}\in\Lambda_{n-1}} \Biggl[ (F^{n-1}\circ W) \left(\sum_{j=1}^{n-1}  \di{j} = \Gamma_{n-1}\right) \\&\cdot \sup_{F_n}(F_{n}\circ W)\left( \di{n} \geq T - \lceil\Gamma_{n-1}\rceil_{\delta}\right)\Biggr]
    \end{split}\\
    \begin{split}
    \leq &\sup_{F^{n-1}}\sum_{\ell=\leastidx}^{\greatestidx} \Biggl[ (F^{n-1}\circ W)\left(\left\lceil\sum_{j=1}^{n-1}  \di{j}\right\rceil_{\delta} = \ell\delta\right) \\&\cdot \sup_{P}\sum_{x}P(x)\sum_y W(y|x) \mathds{1}\left(\left\lceil  \di{n}  + \ell\delta\right\rceil_{\delta}\geq T\right)\Biggr]
    \end{split}\\
    \intertext{Since maximization over $P$ is a linear program over a simplex, the maximum is achieved at a vertex, i.e. the deterministic distribution equal to some $x$.}
    \begin{split}
    =&\sup_{F^{n-1}}\sum_{\ell=\leastidx}^{\greatestidx} \Biggl[ (F^{n-1}\circ W)\left(\left\lceil\sum_{j=1}^{n-1}  \di{j}\right\rceil_{\delta} = \ell\delta\right) \\&\cdot \max_{x}\sum_y W(y|x) \hat{S}^*_n\left(\left\lceil  \di{n}  + \ell\delta\right\rceil_{\delta}\right)\Biggr]
    \end{split}\\
    \begin{split}
    =&\sup_{F^{n-1}}\sum_{\ell=\leastidx}^{\greatestidx} \Biggl[ (F^{n-1}\circ W)\left(\left\lceil\sum_{j=1}^{n-1}  \di{j}\right\rceil_{\delta} = \ell\delta\right){\hat{S}^*_{n-1}(\ell\delta)\Biggr]}
    \end{split}\\
    \intertext{Continuing,}
    =&\sup_{F_1}\sum_{\ell=\leastidx}^{\greatestidx} \left[ (F_1\circ W)\left(\left\lceil \di{1}\right\rceil_{\delta} = \ell\delta\right)\hat{S}^*_{1}(\ell\delta)\right]\\
    =&\,\hat{S}^*_0(0)\,.
\end{align}
\endgroup
\vfill\eject

Thus it suffices to compute $\hat{S}^*_0(0)$, which can be done recursively. Note that
we are not computing the upper bound in~\cite{TimidBold} exactly, but rather an upper
bound on it. The parameter $\delta$ controls the tradeoff between the complexity
of the algorithm and the weakening of the bound.

\section{Algorithms} \label{algorithms}
\subsection{Algorithm for computation of $\hat{S}_0^*(0)$}
\subsubsection{General setup}
The computation of $\hat{S}_0^*(0)$ can be carried out efficiently by defining the transition from one bin to another when using the j-th input as a discrete time-invariant Markov process where the state space is simply the set of all bins. Using this definition we note that the probability of moving from bin $\beta_m$ given that we start at the upper limit of bin $\beta_l$ given the AID change distribution induced by the j-th input only depends on their distance $(l-m)\delta$, and we can thus define this probability as $p_j^{l-m}$. We can now define the Toeplitz transition matrix for the j-th input as
\begin{align}\label{TransitionMatrix}
    M_j=
    \begin{bmatrix}
    p_{j}^{0} & p_j^1 & \cdots & p_j^{N-2}  & p_j^{N-1}\\
    p_{j}^{-1} & p_j^0 & \cdots & p_j^{N-3}  & p_{j}^{N-2}\\
    \vdots & \vdots & \ddots & \vdots & \vdots \\
    p_{j}^{-N+2} & p_j^{-N+3} & \cdots & p_j^0  & p_{j}^{1}\\
    p_j^{-N+1} & p_j^{-N+2} & \cdots & p_j^{-1}  & p_{j}^{0}\\
    \end{bmatrix}\,,
\end{align}
where the max is taken component-wise.
Since by construction the AID can never fall out of the defined range, this stochastic matrix can be used to compute the probability of success vector $\mathbf{\hat{S}}^*_k=[\hat{S}_{k}^*(\leastidx\delta),\cdots,\hat{S}_k^*(\greatestidx\delta)]$ induced by the j-th input for all bins simultaneously as
\begin{gather}
    \mathbf{\hat{S}}^*_{k}=\max_{j\in\{1,\cdots,|\mathcal{X}|\}}M_j\mathbf{\hat{S}}^*_{k+1}\,,
\end{gather}
where the maximum operation is an element-wise maximum over all $M_j\mathbf{\hat{S}}^*_{k+1}$ vectors. We can now easily compute $\hat{S}^*_0(0)$ as shown in algorithm \ref{DP_PS_Optim}.

\begin{algorithm}[hbt!]\label{DP_PS_Optim}
    \caption{Computation of probability of success for optimal controller}
    \KwIn{$W$, $Q$, $n$, $T$, $\delta$}
    \KwOut{$\hat{S}^*_0(0)$}
    Compute $\di{{max}}$ and $\di{{min}}$\\
    Construct bins with spacing $\delta$ in $[n\di{{min}},n\di{{max}}]$\\
    Construct matrices $M_j$ for $j\in\{1,\cdots,|\mathcal{X}|\}$\\
    $\mathbf{\hat{S}}^*_n\leftarrow$ 1 if bin upper bound $>T$ and 0 otherwise\\
    \For{$k\in\{N-1,\cdots,0\}$}{
        $\mathbf{\hat{S}}^*_k=\mathbf{0}$\\
        \For{$j\in\{1,\cdots,|\mathcal{X}|\}$}{
            $\mathbf{\hat{S}}^*_k\leftarrow$ Element-wise max $(\mathbf{\hat{S}}^*_k,M_j\mathbf{\hat{S}}^*_{k+1})$    
        }
    }
    $\hat{S}^*_0(0)\leftarrow$ Bin of $\mathbf{\hat{S}}^*_0$ that contains 0\\
    \textbf{return} $\hat{S}^*_0(0)$
\end{algorithm}

\subsubsection{Algorithm complexity}
The complexity of Algorithm \ref{DP_PS_Optim} depends on the technique used to compute the product $M_j\mathbf{\hat{S}}^*_{k+1}$. One possible method to perform this computation efficiently relies on the fact that all matrices $M_j$ in (\ref{TransitionMatrix}) are Toeplitz, as this fact allows the use of the FFT to speed up computation. In this case the algorithm will have time complexity $O(nN\log(N)|\mathcal{X}|)$ and space complexity $O(N|\mathcal{X}|)$

Another possible method can be derived from the fact that all matrices $M_j$ are generally sparse as when fixing an input the information density change can take at most $|\mathcal{Y}|$ values, which translates to at most $|\mathcal{Y}|$ non-zero elements in each row of the matrix. In this case the algorithm will have time complexity $O(nN|\mathcal{Y}||\mathcal{X}|)$ and space complexity $O(N|\mathcal{X}|)$.


\subsection{Threshold optimization}
\subsubsection{Properties of the optimization landscape}
\begin{figure}[htbp]
  \centering
  \includegraphics[width=0.48\textwidth]{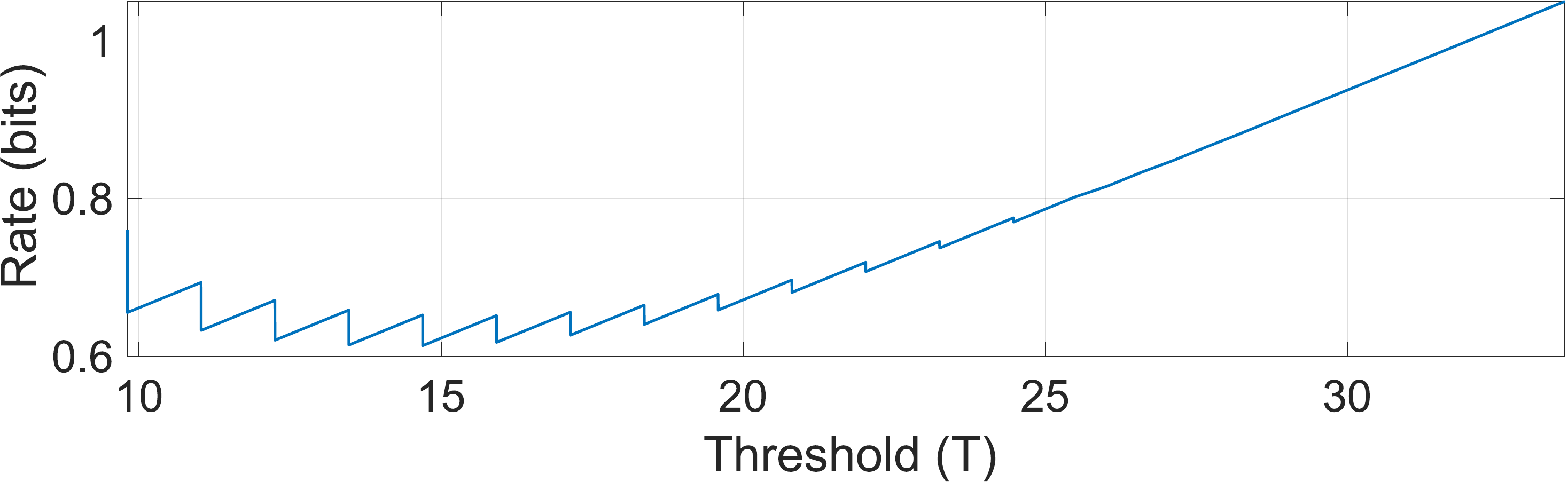}
  \caption{Optimization landscape for BEC with erasure probability $\delta=0.3$ and probability of error $\epsilon=10^{-4}$ at $n=32$}
  \label{fig:BEC_OptLandscape}
\end{figure}

Using algorithm \ref{DP_PS_Optim} we can now obtain a general converse for any DMC by choosing an arbitrary $Q$ that enforces independence between timesteps, and solving the optimization problem derived from equation (\ref{TB_Converse})
\begin{align}\label{ExplicitOptim}
    R^*_b\leq \min_{T}\frac{1}{n}\left[ T-\log_2\left(\left[P(\bar{S}_{F,Q,T})-\epsilon\right]^+\right)\right]\,.
\end{align}
This problem can be difficult to solve using traditional solvers because that the optimization landscape is piecewise linear and discontinuous as shown in Fig \ref{fig:BEC_OptLandscape}. Given a controller $F$ and a threshold $T$ we can express its probability of failure as a sum over all the possible AID outcomes for all controllers using $\Delta I_k$ as defined in (\ref{DeltaIk})
\begin{align}\label{ExplicitPfail}
    P(\bar{S}_{F,Q,T})&=\sum_{\Gamma\in\Lambda_n}\mathds{1}(\Gamma\leq T)\cdot (F\circ W)\left(\sum_{j=1}^{n}\Delta I_{k}=\Gamma\right)
\end{align}
and note that this quantity must be monotonically increasing in $T$ for any arbitrary controller. Therefore, given an optimal controller $F^*$ for a threshold $T$ just above a mass-point $\lambda_m$ if the value of $T$ is increased while staying below the next mass-point with non-zero probability $\lambda_k$ the value of (\ref{ExplicitPfail}) will remain fixed for the original optimal controller as the sum is unchanged and it will still be optimal. 
Thus the bound would grow  linearly until $T$ exceeds $\lambda_k$ where the probability of failure will experience a discrete increase causing the overall bound to potentially drop discontinuously. 

\subsubsection{Heuristics}
It is possible to use the characteristics of the optimization landscape to develop two heuristics that improve the convergence of solvers in certain conditions:
\begin{enumerate}
    \item Our previous discussion shows that a global minimum can only be achieved slightly above a mass-point in $\Lambda$, which makes it possible to restrict the solver to only evaluate the function slightly above valid mass-points.
    \item Intuition and empirical results indicate that the optimal threshold grows approximately linearly with the blocklength. This property allows faster computation when evaluating the converse for an increasing sequence of blocklengths $[n_1,n_2,\cdots]$ as the converses for the first blocklengths can be computed using the full range of thresholds and subsequent converses can be computed by only analyzing a neighborhood around the threshold predicted by a linear regression obtained from the previous values.
\end{enumerate}
Since the converse bound in (\ref{TB_Converse}) is valid for any threshold $T$ the bounds obtained using this process are always valid even if the solver fails to find a minimizing $T$.

\section{Numerical Results and Comparisons} \label{results}

\subsection{Comparison of DP Converse to Existing Converses} \label{sec:BSCBEC}
\begin{figure}[htbp!]
  \centering
  \includegraphics[width=\columnwidth]{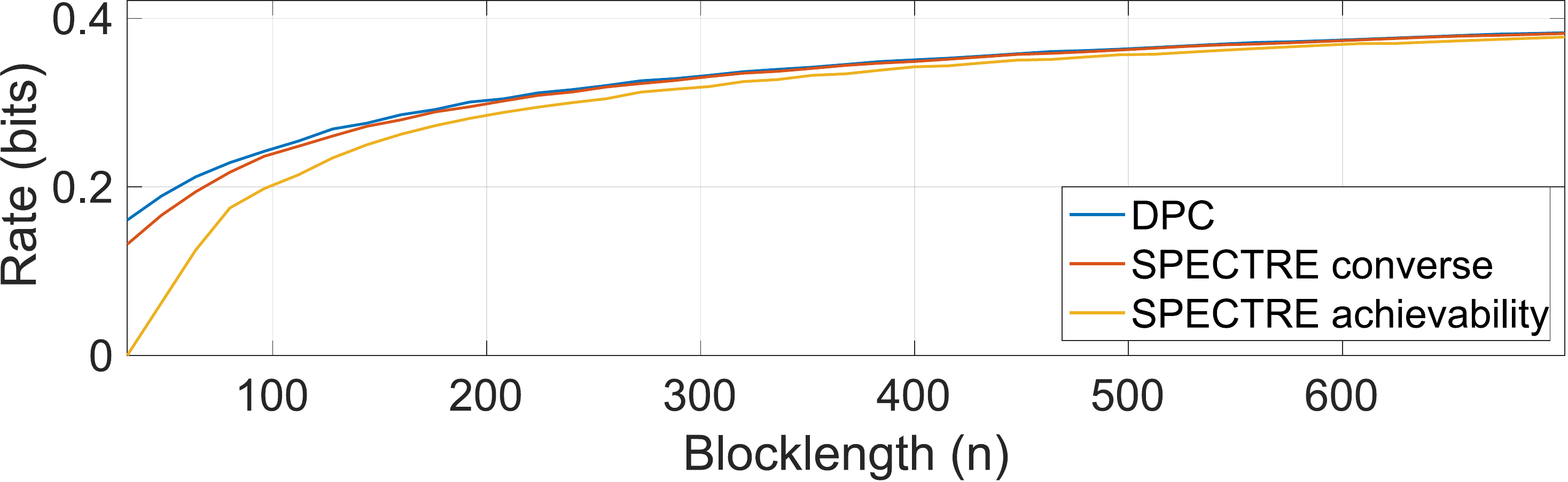}
  \caption{DPC results for BSC with $\epsilon=10^{-4}$ and crossover probability $p=0.11$. The capacity of the channel is approximately 0.5 bits.}
  \label{fig:BSC_Conv}
\end{figure}

\begin{figure}[htbp!]
  \centering
  \includegraphics[width=\columnwidth]{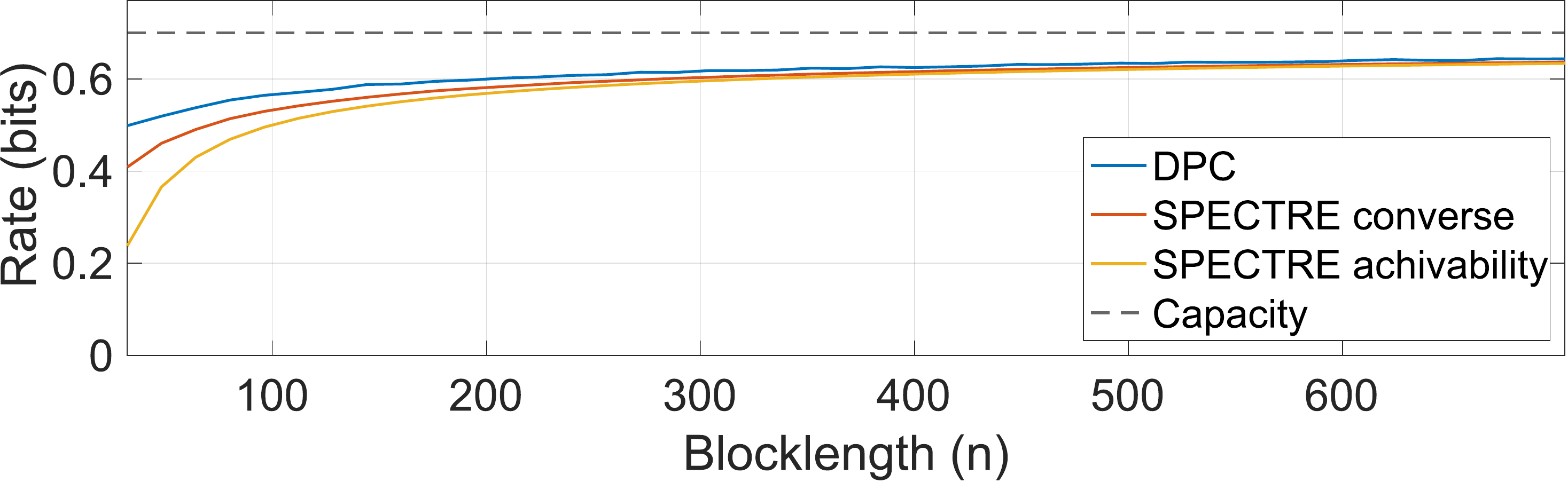}
  \caption{DPC results for BEC with $\epsilon=10^{-4}$ and erasure probability $p_e=0.3$}
  \label{fig:BEC_Conv}
\end{figure}

In this section we compare the results from our DP converse (DPC) to the existing converse and achievability results in the SPECTRE toolbox \cite{spectre} for the BSC and the BEC. These channels are good baselines for the algorithm as they have small input and output alphabets, well known converse and achievability results, and the optimal output distributions are known \cite{polyanskiy_saddlepoint}. In the BSC case our algorithm produces very good results (Fig. \ref{fig:BSC_Conv}) as for blocklengths above 300 the result is less than 1\% above the channel-specific converse, and this gap closes rapidly as the blocklength increases. The BEC case provides a challenge to the DPC algorithm as \cite{polyanskiy_saddlepoint} shows that the optimal distribution is not a product distribution and our algorithm is restricted to product distributions. However, Fig. \ref{fig:BEC_Conv} shows that the DPC result is still very close to the channel specific converse with a gap of less than 2.4\% to the channel specific converse for blocklengths larger than 300. The fact that the optimal output distribution for the BEC is non-product is a possible explanation for the added looseness in the BEC converse when compared to our BSC results, but our algorithm provides good converse results even in this case.



\subsection{New Converse Bounds Using the DP Approach}
\begin{figure}[htbp!]
  \centering
  \includegraphics[width=\columnwidth]{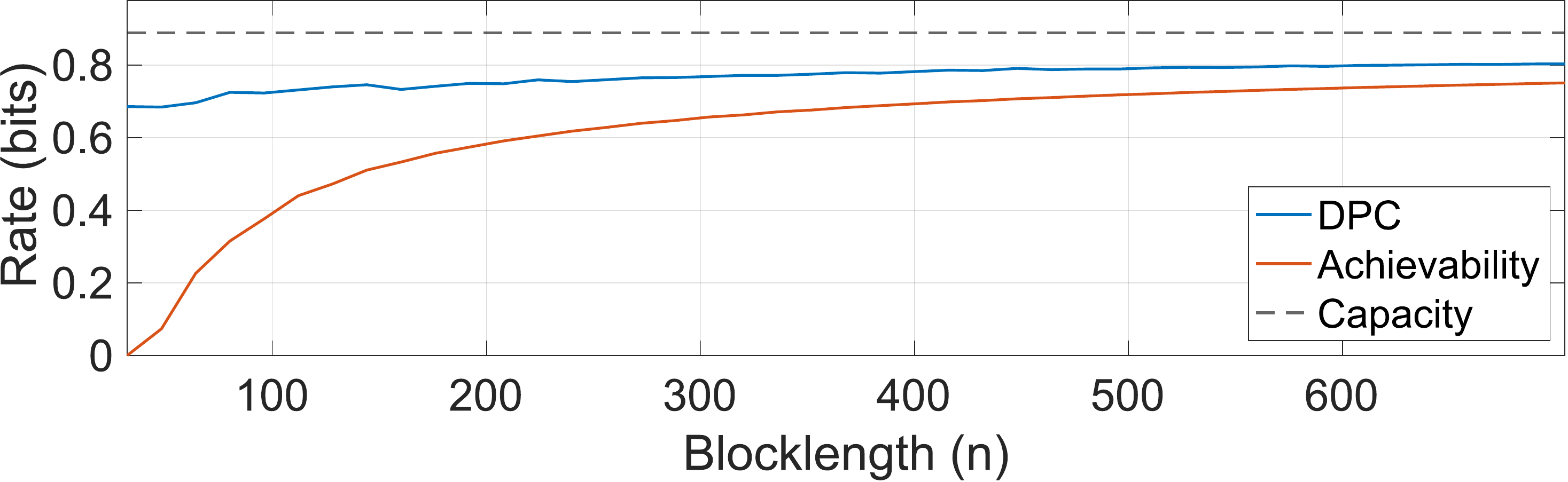}
  \caption{DPC results for BAC with $\epsilon=10^{-4}$ and crossover probabilities $p_1=0.01$, $p_2=0.02$. The achievability curve is derived from using Lemma 14 in \cite{TimidBold} with a capacity achieving distribution.}
  \label{fig:BAC_Conv}
\end{figure}

Section \ref{sec:BSCBEC}  shows that the performance of our DP algorithm is comparable to channel specific converses.  This section applies the new algorithm to two DMCs not covered by the SPECTRE toolbox. Fig. \ref{fig:BAC_Conv} shows that the DPC produces a reasonable converse for the BAC. For blocklengths above 500 the result is less than 9.9\% above the achievable rate computed using Lemma 14 of \cite{TimidBold} with a capacity-achieving distribution. It is hard to determine whether this gap originates from looseness of the achievability or converse results, but the behavior of our converse is consistent with the BSC and BEC results.

\begin{figure}[htbp!]
    \centering
    \includegraphics[width=\columnwidth]{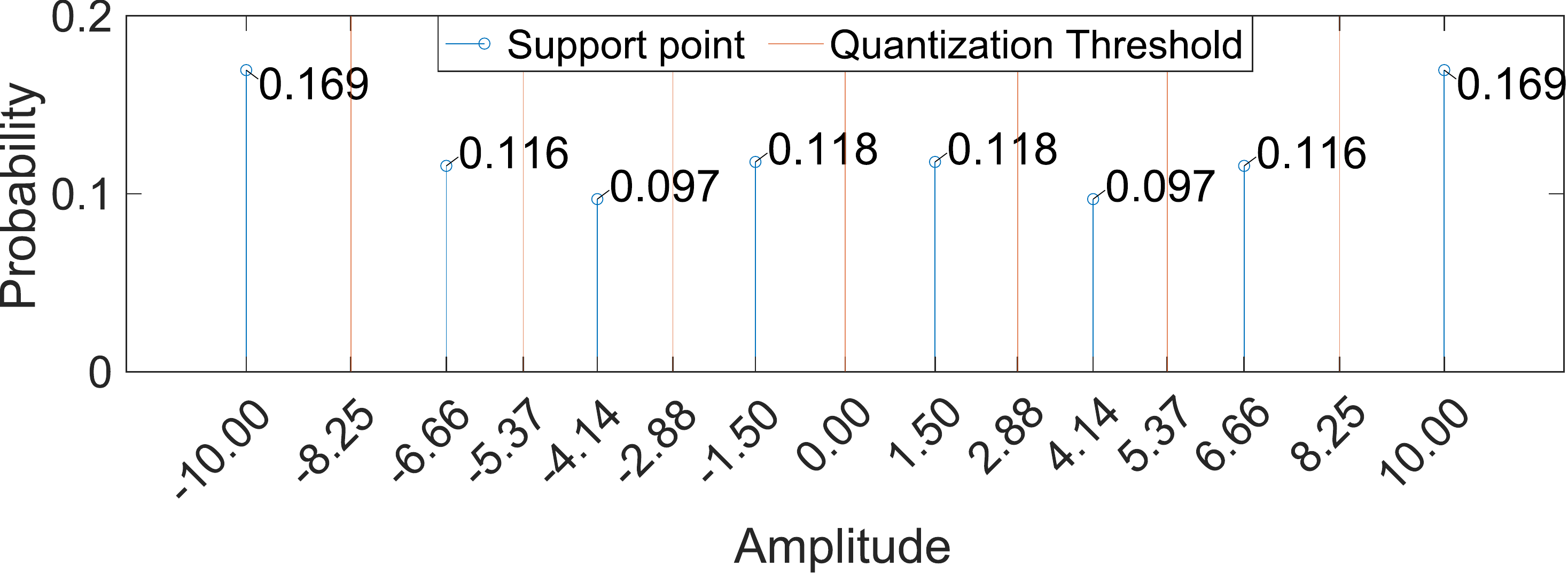}
    \caption{Constellation and quantization thresholds that maximize mutual information for the unit noise AWGN channel with quantized output under amplitude constraint 10. These induce the DMC given by $W$ shown above.}
    \label{fig:DABQ_const}
\end{figure}

Recall that existing algorithms for computing converses have complexity that scales with
the blocklength $n$ and input alphabet size $|\mathcal{X}|$ as $n^{|\mathcal{X}|}$.
Next we show that our DP algorithm can be used to provide bounds at moderately large 
$n$ and $\mathcal{X}$.
Consider a DMC with $|\mathcal{X}| = |\mathcal{Y}| = 8$ corresponding to using an AWGN channel with an optimized input alphabet and quantized outputs as described in \cite{madhow_awgnq}.  The input alphabet is a fixed set of constellation points $\mathcal{X}$.  The output is quantized using thresholds $-\infty = q_0 <q_1<q_2<\dots<q_7<q_8=\infty$.
With input $x$, the output is $i$ if $x+Z \in (q_{i-1}, q_i]$, where $Z\sim \mathcal{N}(0,1)$.
Such channels are of practical interest, and it is known that the optimal input distribution has finite-support \cite{madhow_awgnq} given some fixed quantization bins $\{q_i\}$. Dynamic Assignment Blahut-Arimoto \cite{dab_awgn, dab_pic} can be used to find the optimal $\mathcal{X}$ and input distribution when restricting the input of a channel to be finite-support.
Modifying the  algorithm for AWGN channel with amplitude constraint from \cite{dab_awgn} to account for output quantization and performing alternating optimization with $\{q_i\}$ gives a choice of $\mathcal{X}$ and $\{q_i\}$.
Setting $|\mathcal{Y}| = 8$ and the amplitude constraint to $10$ yields an optimized $\mathcal{X}$ of cardinality $8$ as shown in Fig. \ref{fig:DABQ_const}, which yields the DMC described by transitions matrix $W$ shown in Fig.~\ref{fig:BigChannel}.

\begin{figure}[htbp]
\begin{align*}
\label{eq:dabq_W}
\small
W \approx 
\begin{bmatrix}
0.96& 0.04& 0.00& 0.00& 0.00& 0.00& 0.00& 0.00\\
0.06& 0.84& 0.10& 0.00& 0.00& 0.00& 0.00& 0.00\\
0.00& 0.11& 0.79& 0.10& 0.00& 0.00& 0.00& 0.00\\
0.00& 0.00& 0.08& 0.85& 0.07& 0.00& 0.00& 0.00\\
0.00& 0.00& 0.00& 0.07& 0.85& 0.08& 0.00& 0.00\\
0.00& 0.00& 0.00& 0.00& 0.10& 0.79& 0.11& 0.00\\
0.00& 0.00& 0.00& 0.00& 0.00& 0.10& 0.84& 0.06\\
0.00& 0.00& 0.00& 0.00& 0.00& 0.00& 0.04& 0.96
\end{bmatrix}
\end{align*}
    \caption{Equivalent DMC representation for the quantized output amplitude constrained AWGN channel.}
  \label{fig:BigChannel}
\end{figure}
\begin{figure}[htbp!]
  \centering
  \includegraphics[width=\columnwidth]{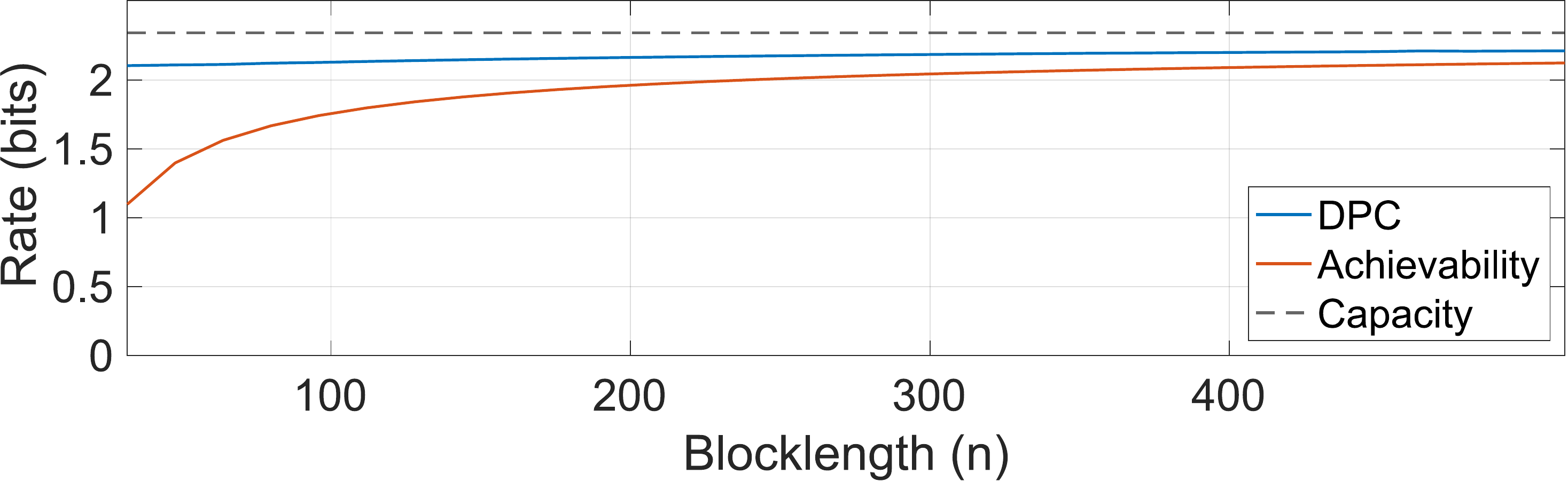}
  \caption{DPC results for finite-alphabet, quantized-output AWGN channel described by the $W$ shown above with $\epsilon=10^{-4}$. The achievability curve is derived using Lemma 14 in \cite{TimidBold} with a capacity achieving distribution.}
  \label{fig:DABQ_Conv}
\end{figure}

Fig. \ref{fig:DABQ_Conv} shows that the DPC also provides a reasonable converse in this case.  For blocklengths above 300 the converse is less than 6.8\% above the achievable rate computed using Lemma 14 of \cite{TimidBold} with a capacity-achieving distribution.

\section{Conclusion} \label{conclusion}
This paper presents a general technique for computing finite-blocklength converse bounds for an arbitrary DMC.  The converse uses dynamic programming to solve a stochastic control formulation of the converse problem.  The utility of the approach was verified for the BSC, BEC, BAC and a DMC resulting from an amplitude constrained AWGN channel with quantized output.  Since the technique presented provides a valid bound for any product output distribution, the bound can be further improved by optimizing the output distribution over this set. Identifying the best such output distribution is a topic for future work, as is the extension to channels with cost constraints.
\newpage
\bibliographystyle{IEEEtran}
\bibliography{Ref-wagner}

\begin{thebibliography}{10}
\providecommand{\url}[1]{#1}
\csname url@samestyle\endcsname
\providecommand{\newblock}{\relax}
\providecommand{\bibinfo}[2]{#2}
\providecommand{\BIBentrySTDinterwordspacing}{\spaceskip=0pt\relax}
\providecommand{\BIBentryALTinterwordstretchfactor}{4}
\providecommand{\BIBentryALTinterwordspacing}{\spaceskip=\fontdimen2\font plus
\BIBentryALTinterwordstretchfactor\fontdimen3\font minus
  \fontdimen4\font\relax}
\providecommand{\BIBforeignlanguage}[2]{{%
\expandafter\ifx\csname l@#1\endcsname\relax
\typeout{** WARNING: IEEEtran.bst: No hyphenation pattern has been}%
\typeout{** loaded for the language `#1'. Using the pattern for}%
\typeout{** the default language instead.}%
\else
\language=\csname l@#1\endcsname
\fi
#2}}
\providecommand{\BIBdecl}{\relax}
\BIBdecl

\bibitem{Blahut}
R.~E. Blahut, ``Hypothesis testing and information theory,'' \emph{IEEE
  Transactions on Information Theory}, vol.~20, no.~4, pp. 405--417, 1974.

\bibitem{Altug:SP:Asymmetric}
Y.~Altu\u{g} and A.~B. Wagner, ``Refinement of the sphere-packing bound:
  Asymmetric channels,'' \emph{IEEE Trans.\ Inf.\ Theory}, vol.~60, no.~3, pp.
  1592--1614, Mar. 2014.

\bibitem{Altug:EA:Symmetric}
------, ``On exact asymptotics of the error probability in channel coding:
  symmetric channels,'' \emph{IEEE Trans.\ Inf.\ Theory}, vol.~67, no.~2, pp.
  844--868, Feb. 2021.

\bibitem{ppv_2010}
Y.~Polyanskiy, H.~V. Poor, and S.~Verdu, ``Channel coding rate in the finite
  blocklength regime,'' \emph{IEEE Transactions on Information Theory},
  vol.~56, no.~5, pp. 2307--2359, 2010.

\bibitem{polyanskiy_saddlepoint}
Y.~Polyanskiy, ``Saddle point in the minimax converse for channel coding,''
  \emph{IEEE Transactions on Information Theory}, vol.~59, no.~5, pp.
  2576--2595, 2013.

\bibitem{ErsegheAWGN}
\BIBentryALTinterwordspacing
T.~Erseghe, ``On the evaluation of the polyanskiy-poor-verdu converse bound for
  finite blocklength coding in {AWGN},'' \emph{CoRR}, vol. abs/1401.7169, 2014.
  [Online]. Available: \url{http://arxiv.org/abs/1401.7169}
\BIBentrySTDinterwordspacing

\bibitem{elkayam2016calculation}
N.~Elkayam and M.~Feder, ``On the calculation of the minimax-converse of the
  channel coding problem,'' 2016.

\bibitem{matthews_linear_2012}
W.~Matthews, ``A linear program for the finite block length converse of
  {Polyanskiy–Poor–Verdú} via nonsignaling codes,'' \emph{IEEE
  Transactions on Information Theory}, vol.~58, no.~12, pp. 7036--7044, 2012.

\bibitem{kosut_boosting_2015}
O.~Kosut, ``Boosting output distributions in finite blocklength channel coding
  converse bounds,'' in \emph{2015 IEEE Information Theory Workshop (ITW)},
  2015, pp. 1--5.

\bibitem{vilar2018saddlepoint}
G.~Vazquez-Vilar, A.~G. i~Fabregas, T.~Koch, and A.~Lancho, ``Saddlepoint
  approximation of the error probability of binary hypothesis testing,'' in
  \emph{2018 IEEE International Symposium on Information Theory (ISIT)}, 2018,
  pp. 2306--2310.

\bibitem{ErsegheFull}
T.~Erseghe, ``Coding in the finite-blocklength regime: Bounds based on laplace
  integrals and their asymptotic approximations,'' \emph{IEEE Transactions on
  Information Theory}, vol.~62, no.~12, pp. 6854--6883, 2016.

\bibitem{tatikonda_capacity_2009}
S.~Tatikonda and S.~Mitter, ``The capacity of channels with feedback,''
  \emph{IEEE Transactions on Information Theory}, vol.~55, no.~1, pp. 323--349,
  2009.

\bibitem{TimidBold}
A.~B. Wagner, N.~V. Shende, and Y.~Altuğ, ``A new method for employing
  feedback to improve coding performance,'' \emph{IEEE Transactions on
  Information Theory}, vol.~66, no.~11, pp. 6660--6681, 2020.

\bibitem{tatikonda_control_2000}
S.~C. Tatikonda, ``Control under communication constraints,'' Ph.D.
  dissertation, Electrical Engineering and Computer Science, Massachusetts
  Institute of Technology, Cambridge, MA, 8 2000.

\bibitem{spectre}
\BIBentryALTinterwordspacing
S.~Chen, A.~Collins, G.~Durisi, T.~Erseghe, G.~C. Ferrante, V.~Kostina,
  J.~Östman, Y.~Polyanskiy, I.~Tal, and W.~Yang, ``{SPECTRE}: short-packet
  communication toolbox,'' 2021. [Online]. Available:
  \url{github.com/yp-mit/spectre}
\BIBentrySTDinterwordspacing

\bibitem{madhow_awgnq}
J.~Singh, O.~Dabeer, and U.~Madhow, ``On the limits of communication with
  low-precision analog-to-digital conversion at the receiver,'' \emph{IEEE
  Transactions on Communications}, vol.~57, no.~12, pp. 3629--3639, 2009.

\bibitem{dab_awgn}
D.~Xiao, L.~Wang, D.~Song, and R.~D. Wesel, ``Finite-support
  capacity-approaching distributions for {AWGN} channels,'' in \emph{2020 IEEE
  Information Theory Workshop (ITW)}, 2021, pp. 1--5.

\bibitem{dab_pic}
N.~Farsad, W.~Chuang, A.~Goldsmith, C.~Komninakis, M.~Médard, C.~Rose,
  L.~Vandenberghe, E.~E. Wesel, and R.~D. Wesel, ``Capacities and optimal input
  distributions for particle-intensity channels,'' \emph{IEEE Transactions on
  Molecular, Biological and Multi-Scale Communications}, vol.~6, no.~3, pp.
  220--232, 2020.

\end{thebibliography}

\end{document}